\documentclass[showpacs,preprintnumbers,amssymb]{revtex4}

\usepackage{graphicx}
\usepackage{dcolumn}
\usepackage{bm}
\usepackage{epsf}

\newcommand{\beq}{\begin{equation}}
\newcommand{\eeq}{\end{equation}}
\newcommand\beqa{\begin{eqnarray}}
\newcommand\eeqa{\end{eqnarray}}
\newcommand\bea{\begin{array}}
\newcommand\eea{\end{array}}
\newcommand\ba{\begin{array}}
\newcommand\ea{\end{array}}
\newcommand{\nn}{\nonumber}

\newcommand{\neqa}{\nonumber\end{eqnarray}}
\newcommand{\la}{\label}

\newcommand{\eq}[1]{eq.(\ref{#1})}

\renewcommand{\d}{{\partial}}

\newcommand{\bastar}{\begin{eqnarray*}}
\newcommand{\eastar}{\end{eqnarray*}}

\def\D{{\nabla}}
\def\nn{\nonumber}

\def\tP{{\tilde{P}}}
\def\C{C}
\def\r{{\varrho}}
\def\scip{\;;\;\;}
\newcommand{\ce}{{\mathbb e}}

\begin{document}

\title{Yang-Mills instanton as a quantum black hole}

\begin{abstract}
In terms of spin-charge separated variables, the Minkowski space
Yang-Mills BPST instanton describes a locally conformally flat
doubly-wrapped cigar manifold
that can be viewed as a Euclidean quantum black hole.
An ensemble of instantons then corresponds to a ``spacetime foam'' that
creates a locally conformally flat spacetime from ``nothing'' as a quantum
fluctuation.
\end{abstract}

\author{Sergey Slizovskiy}

\email{Sergey.Slizovskiy@teorfys.uu.se} \vskip 0.3true cm

\affiliation{ Department of Theoretical Physics,
Uppsala University,
P.O. Box 803, S-75108, Uppsala, Sweden \\
St.Petersburg INP, Gatchina, 188 300, St.Petersburg, Russia}

\maketitle
\section{Motivation}
One of the challenges in quantum gravity is to find a
background independent formulation. The classical spacetime
should emerge only as a low energy approximation to a more fundamental
and yet unknown description of quantum spacetime.

\vskip 0.3cm

In a recent article Faddeev and Niemi \cite{FaddNiemi} mapped
the four dimensional SU(2) Yang-Mills theory into Einstein's gravitational
theory with matter, and with conformal geometry. Their mapping is
a change of variables that can be interpreted in terms of a
separation between the spin and the charge of the gauge field.
Their result shows that in the Maximal Abelian Gauge (MAG) the Yang-Mills theory
admits a generally covariant form that includes the Einstein-Hilbert action and
a cosmological constant term, in the limit where the metric tensor
is locally conformally flat. The matter content is a combination of a
massive vector field, an $O(3)$ $\sigma$-model and a $G(4,2)$ Grassmannian
nonlinear $\sigma$-model that describes the embedding of two dimensional planes
(2-branes) in $\mathbb R^4$.

The algebraic change of variables by Faddeev and Niemi can be performed either
in the full effective action of the quantum Yang-Mills theory, or in its
low energy limit. In the former case we obtain - by construction - a renormalizable
quantum theory of gravity with matter. This theory could be interpreted as a subsector
of some more realistic quantum gravity theory. In the latter case where we only include
the leading terms in the effective action corresponding to the classical Yang-Mills
action and a potential term for a dimension-two condensate that appears in the
Maximal Abelian Gauge  \cite{FaddNiemi}, the conformal scale of the metric tensor
turns out to coincide with the
dimension-two condensate introduced in \cite{Gubarev2000,Gubarev,Verschelde,Kondo2001}.
As a consequence it becomes meaningful to address the question, what kind of quantum
fluctuations give rise to a non-zero metric tensor condensate?

A partial answer to this question can be inferred from \cite{Brasil,CondensateReview}: The
dimension-two condensate of \cite{Gubarev2000,Gubarev,Verschelde,Kondo2001}
is known to receive a nontrivial contribution from perturbative quantum fluctuations around a
homogeneous background.

In the present article we are interested in those non-perturbative contributions
to the metric tensor condensate that emerge from the BPST instanton solutions.
Indeed, it has been recently argued that the dimension-two condensate of
\cite{Gubarev2000,Gubarev,Verschelde,Kondo2001} should receive a significant contribution
from instantons \cite{Boucaud} \footnote{Although
in \cite{Boucaud} the condensate is estimated in the Landau gauge, we expect
qualitatively the results to persist in the MAG gauge}.
This suggests that when the BPST instanton is described  in the geometric formalism by Faddeev
and Niemi,  it serves as a source of non-perturbative fluctuations
that give rise to a non-trivial locally conformally flat spacetime.




In section \ref{sec2} we will briefly review the formalism of spin-charge separation.
In sections \ref{sec3},\ref{sec4},\ref{sec5} we study several aspects of instanton solutions in
this formalism, and our focus is on a geometrical interpretation. The results
are summarized in the final section \ref{secDiscussion}.

\section{Spin-charge separation} \la{sec2}
We start with a short description of the spin-charge
decomposition  for the $SU(2)$
Yang-Mills action \footnote{We also expect that some
other types of decompositions of Yang-Mills variables could
have have gravitational interpretation \cite{Maxim}}; we refer to  \cite{FaddNiemi} for details.
We decompose the $su(2)$ vector-potential as
\beqa
\hat A_\mu = A_\mu \frac{\sigma^3}2 + X_{\mu +} \frac{\sigma^-}2 + X_{\mu -} \frac{\sigma^+}2
\eeqa
where we have denoted
\beqa
\sigma_\pm  = \frac12 (\sigma_1 \pm i \sigma_2)
\\
 X_{\mu\pm} = A^1_\mu \pm i A^2_\mu
\eeqa

We fix the $SU(2)/U(1)$ gauge freedom by employing the Maximal Abelian Gauge (MAG) condition,
locally defined as
\beq \la{MAG}
 \nabla_\mu^{\pm} X_{\mu \pm} = 0
\eeq
where the Abelian covariant derivative $\nabla$ is
\beq
\D^{\pm}_\mu = \d_\mu \pm i A_\mu
\eeq
In order to avoid Gribov copies, we also demand that the gauge fields
satisfy the global MAG condition
\beq \la{GlobalMAG}
 A: \; \int \left({\left.^\Omega \!\!A_\mu^1 \right.}\right)^2
+ \left({\left.^\Omega \!\!A_\mu^2 \right.}\right)^2  d^4 x \ \ \ \mbox{ \ \ is \ minimal
\ at }\Omega=1
\eeq
Here $\left.^\Omega \!\!A_\mu \right.$ is the gauge transformation of $A_\mu$.
The remaining $U(1) \in SU(2) $ gauge freedom will not be gauge fixed. Instead,
we will eliminate this remaining gauge freedom by introducing
explicitly  $U(1)$-invariant variables.

We note \cite{FaddNiemi} that the direction of the
$U(1)$ Cartan subalgebra in the $SU(2)$ Lie algebra can be chosen arbitrarily.
However, for simplicity we here proceed with a
Cartan subalgebra that coincides with
the $\sigma^3$ generator of the SU(2) Lie algebra.

Following \cite{FaddNiemi} we introduce the spin and the charge variables:
\beq \ba{c}
 X_{\mu+} = \psi_1 e_\mu + \psi_2 e_\mu^*\\
 X_{\mu-} =  \psi_2^* e_\mu + \psi_1^* e_\mu^*
 \ea
 \la{defpsi}
\eeq
where the complex field $e_\mu$ is a complex combination
of zweibeine
\beq
e_\mu = \frac{e_\mu^1 + i e_\mu^2}{\sqrt2}
\eeq
 normalized as
\beq
\begin{array}{c} e_\mu e_\mu=0 \\
 e_\mu e_\mu^*=1 .\end{array}
\eeq
The internal $U(1)_I$ symmetry of the decomposition is defined by the transformations
\beqa
 \psi_1 \to e^{i \lambda} \psi_1 \\
  \psi_2 \to e^{-i \lambda} \psi_2 \\
  \phi \to \phi - 2 \lambda \\
e_\mu \to e_\mu \,e^{-i \lambda}
\eeqa

We can factor out the $U(1)_I$ phase variable as follows:
If $e_0 \neq 0$, we define $\eta$ to be the
complex phase of $e_0$ and
\beq
\hat e_\mu = e^{-i \eta} e_\mu,
\la{ehat}
\la{defeta}
\eeq
if $e_0=0$, we can use other charts to define $\eta$, so it is a section of $U(1)_I$ bundle over ${\mathbb R}^4$.

The variable $\rho$ is defined as,
\beq
\la{def_rho}
\rho^2=(A_\mu^1)^2 + (A_\mu^2)^2 = X_{\mu+} X_{\mu -}
\eeq
It corresponds to the gauge invariant minimum of the
functional (\ref{GlobalMAG}) prior to any gauge fixing.
In the framework of effective low-energy theory, $\rho$
is the dimension-two condensate of \cite{Gubarev2000,Gubarev,Verschelde,Kondo2001}.

An explicit parameterization for $\psi_{1,2}$ in \eq{defpsi} is given by
\beqa
\psi_1 = \rho\, e^{i \xi}
\cos\!\frac\theta 2 \,e^{-i \phi/2} \nn
 \\
\psi_2 = \rho\, e^{i \xi} \sin\!\frac\theta 2 \,e^{i \phi/2}.
\la{psi_param}
\eeqa

We introduce the $U(1)_I$ gauge field as a Maurer-Cartan 1-form:
\beq
\la{defC}
\C_\mu = i <e^*,\d_\mu e>.
\eeq



Define also the $U(1)_I$ invariant gauge field
\beq
 \hat C_\mu = i <\hat e^*,\d_\mu \hat e>=\C_\mu + \d_\mu \eta
\eeq
and the $U(1)\times U(1)_I$ invariant 3-vector
\beq
n = \left(\begin{array}{c}
  \sin\theta \cos(\phi- 2 \eta) \\
  \sin\theta \sin(\phi- 2 \eta) \\
  \cos\theta
\end{array}\right)
\la{defn}
\eeq
\beq  \la{n_pm}
  n_\pm = n_1 \pm i n_2  = \sin \theta e^{\pm i (\phi-2 \eta)}.
\eeq
The tensor $P_{\mu\nu}$ is
\beq
  P_{\mu\nu} = \frac12 \rho^2 n_3  \tP_{\mu\nu}
\eeq
with $\tP$ given by
\beq
 \tP_{\mu\nu} = i (e_\mu e^*_\nu - e_\nu e^*_\mu).
\eeq

A doublet of $U(1)\times U(1)_I$ invariant, mutually orthogonal and
unit normalized 3-vectors $p$ and $q$ can also be used in lieu of the
zweibeine:
\beqa p_i=\frac12 \tP_{i 0} \;\;;\;\;
q_i = \frac14 \epsilon_{ijk} \tP_{jk}
\la{defpq}
\eeqa
\beq
\ba{ccc}
p \cdot q = 0 & ; & p \cdot p +q \cdot q = \frac14\\
p \cdot p = \frac12 e_0 e_0^* &  ;  &
q \cdot q = \frac14 - \frac12 e_0 e_0^*.
\ea
\eeq

Finally, we define the combined $U(1) \times U(1)_I$ covariant derivatives
\beqa
 D_\mu \psi_1 = \d_\mu \psi_1 + i A_\mu \psi_1 -i \C_\mu \psi_1\\
 D_\mu \psi_2 = \d_\mu \psi_2 + i A_\mu \psi_2 + i \C_\mu \psi_2\\
 D_\mu e_\nu = \d_\mu e_\nu + i \C_\mu e_\nu
\eeqa and the $U(1) \times U(1)_I$ invariant current
\beq
 J_{\mu} = \frac i{2 \rho^2} \{\psi_1^*D_\mu \psi_1 -\psi_1 \bar D_\mu \psi_1^*
 +\psi_2^* D_\mu \psi_2 -\psi_2 \bar D_\mu \psi_2^*\}
\eeq

In terms of these variables the standard classical Yang-Mills Lagrangian with a
gauge fixing term for the off-diagonal components is
\cite{FaddNiemi}:
\beqa
 L&=&\frac14 (F^i_{\mu \nu})^2+\frac{\xi}{2} |\D^+_\mu X_{\mu+}|^2  \\ &=&
  \frac{1}{4}{\mathcal F}_{\mu\nu}^2
+ \frac{1}{2}(\partial_\mu \rho)^2
+ \frac{1}{2}\rho^2  J_\mu^2 + \frac{1}{8}\rho^2  (\mathrm
D^{\hat C}_{\, \mu}
\vec n)^2
+ \rho^2 \left((\d_\mu q)^2 + (\d_\mu p)^2 \right) \nn\\&&
+ \frac{1}{4} \rho^2
\left\{ n^{}_+ ( \partial_a\hat{\bar e}_b)^2 + n^{}_-
(\partial_a \hat e_b)^2 \right\}
+ \frac{3}{8} (1-n_3^2) \rho^4 - \frac{3}{8} \rho^4 -  \frac12 \d_\nu\d_\mu (X_{\mu +} X_{\nu -})
\la{YMfi}
\eeqa
where in the second equality we have taken the $\xi \to \infty$ limit corresponding
to the MAG gauge condition (\ref{MAG}) and
\beqa
\mathcal F_{\mu \nu} =  \partial_\mu J_\nu - \partial_\nu J_\mu
+ \frac{1}{2} \, \vec n \cdot \partial_{\mu} \vec n \times
\partial_{\nu} \vec n -  \{ \partial_\mu (n_3\hat C_\nu) -
\partial_\nu (n_3 \hat C_\mu) \} -
\rho^2 n_3 \tilde P_{\mu\nu}
\\
(\mathrm D^{\hat C}_{\, a})^{ij} = \delta^{ij} \partial_a +
2 \epsilon^{ij3} \hat C_a
\ \ \ \ \ \ \ (i,j=1,2,3)
\la{nabladef}
\eeqa


\section{One instanton configuration in spin-charge separated variables} \la{sec3}
We recall the explicit BPST one-(anti)instanton solution
of the $SU(2)$ gauge theory \cite{BPST}, in the singular gauge it reads
\beq \la{inst}
 A_\mu^a(x) = \r^2 \frac{2\, \eta_{a\mu\nu} x^\nu}{x^2 (x^2+\r^2)}.
\eeq
where $\eta_{aij}= \epsilon_{aij}$ \ ; \ $\eta_{a4\nu}= -\eta_{a\nu 4}= -i\delta_{a\nu}$
is the 't Hooft tensor \cite{tHooft} and for simplicity we have centered the instanton at the origin.
It is widely known and can be easily verified, that this explicit representation
obeys the MAG conditions (\ref{MAG}).
In particular, the explicit solution (\ref{inst}) is known to be the {\it global}
gauge orbit minimum of the MAG gauge fixing functional
\[
\int d^4 x \rho^2
\]
thus it also obeys the global MAG condition (\ref{GlobalMAG}).
We note that this configuration lies on the Gribov \cite{Gribov} horizon, and the explicit ghost
zero modes have been  constructed in \cite{Bruckmann2000}.
Thus (\ref{inst}) is a singular point in the field space,
where the fundamental modular region reaches the Gribov horizon.



The instanton solution (\ref{inst}) depends on five parameters,
the instanton size $\r$ and the four components $X_m$ of the vector that defines
the instanton position. These leave the MAG condition intact.
Furthermore, when one introduces global
$SU(2)$ gauge-rotations in the right hand side of \eq{inst}
the total number of free parameters of a single-instanton solution becomes eight.
Since a global gauge rotation is equivalent to a space-time rotation of the solution,
the global gauge rotation also does not violate the MAG condition.

Notice that in general one could expect that the MAG condition fixes
the global $SU(2)/U(1)$ gauge freedom. However, for instanton it does not, which also
indicates that the instanton is located on the Gribov horizon.
We also note that in \cite{Brower}  a continuous family
of gauge transformations, smoothly interpolating
between regular and singular gauges and satisfying the differential version
of the MAG gauge condition has been presented. This suggests that
all of them should comprise a valley of degenerate classical solutions for the resulting
action. But here we only consider the singular-gauge solution since only this solution
lies in the fundamental modular domain.


We now proceed to rewrite the instanton solution
in  the spin-charge separated variables.
We use the relation \cite{tHooft}
\[
\eta_{a\mu\nu} \eta_{b\mu\lambda} =
\delta_{ab} \delta_{\nu\lambda} + \varepsilon_{abc} \eta_{c \nu \lambda}
\]
to get
\beq \la{rho_inst}
 \rho^2 =  \frac{8 \r^4}{(\r^2+x^2)^2 x^2}
\eeq

The zweibein $e_\mu^a(x)$ defines the 2-plane
in which $A^a$ lies:
\beqa
e^a_\mu &=&
\,\eta_{a\mu\nu} \frac {x^\nu}{x}.
\la{einst}
\eeqa
This choice corresponds to a particular fixing of the $U(1)_I$ gauge freedom:
The $U(1)_{I}$ phase $\eta$ which is defined in (\ref{defeta}), is now
constant. Hence it disappears entirely. Note that $e^a_\mu$ is ill-defined in
the origin but we shall find that the underlying
geometry removes this point.

When we choose a gauge where $\psi_{1,2}$ are both real
(this fixes both the $U(1)$ and the $U(1)_{I}$ gauges)
we have by definition that
\[
A_\mu^1 = (\psi_1+\psi_2) e^1_\mu \\
A_\mu^2 = (\psi_1-\psi_2) e^2_\mu
\]
So we find
\beq
\psi_{1} = \rho \;\; \ \; \& \ \ \  \psi_2 = 0 \;\; \Rightarrow \;\;\ \theta = 0,
\eeq
which means that the 3-vector $\vec n$, defined in (\ref{defn}),
points identically to its third (vacuum) direction. Thus this vector field
has no classical dynamics. In particular, this configuration
is manifestly Lorentz invariant \cite{FaddNiemi}.

Generally,  $n_3 = \pm 1$ is achieved when
$(A^1_\mu)^2 = (A^2_\mu)^2$, and a nontrivial $\vec n$ accounts for deviations
from this equality.

From (\ref{defpq}) we calculate the vectors $p$ and $q$:
\beqa
p_i=\frac{x_\mu x_\nu}{2 x^2} (\eta_{1i\nu} \eta_{24\mu} -\eta_{2i\nu} \eta_{14\mu}) \\
q_i= \frac{x_\mu x_\nu}{2 x^2} \epsilon^{i j k} \eta_{1j\mu} \eta_{2k\nu}
\eeqa
The internal connection (the one characterizing the $U(1)_I$ gauge bundle)
defined in (\ref{defC}), is
\beq
C_\mu =\eta_{3 \mu\nu} \frac{x^\nu}{x^2} =  \left\{\frac{x_2}{x^2},-\frac{x_1}{x^2},\frac{x_4}{x^2},-\frac{x_3}{x^2}\right\}.
\eeq
The $U(1) \times U(1)_I$ invariant current is simply
\beq
 J_\mu = C_\mu - A_\mu^3.
\eeq

Since $\vec n$ is trivial, we conclude that the BPST instanton
is also a classical  solution to the equations of motion for the {\it restricted}
Yang-Mills Lagrangian
\beqa
\la{Lreduced}
 L&=&
  \frac{1}{4}{\mathcal F}_{ab}^2
+ \frac{1}{2}(\partial_a \rho)^2
+ \frac{1}{2}\rho^2  J_a^2
+ \rho^2 \left((\d_\mu p)^2 + (\d_\mu q)^2 \right)
 - \frac{3}{8} \rho^4 -  \frac12 \d_\nu\d_\mu (X_{\mu +} X_{\nu -})
\la{YMreduced}
\eeqa
where
\beqa
\mathcal F_{ab} &=&  \partial_a J_b - \partial_b J_a
 -  \{ \partial_a (\hat C_b) -
\partial_b (\hat C_a) \} -
\rho^2 \tilde P_{ab} \nn \\&=& -\partial_a A_b + \partial_b A_a -
\rho^2 \tilde P_{ab}.
\eeqa

Note that for the anti-instanton we find the same solution but
with 't Hooft tensor$\eta \to \bar\eta$. To account for arbitrary color orientation
of (anti-)instanton, one has to rotate all spacetime indices by the $SU(2)$ matrix
acting as  $SU(2)_{L \,(R)}$ subgroup of $SO(4)$ or Lorentz group.

We now turn to the generally covariant formulation of \cite{FaddNiemi}. There, it is
proposed that $\rho$ can be viewed as the conformal scale of a conformally flat metric tensor,
\beq \la{gdef}
 g = 1_{4 \! \times 4} \, (\rho/\Delta)^2 ,
\eeq
where $\Delta$ is a constant with the dimension of mass which we need to introduce since
$\rho$ is dimension-two operator while the metric tensor is dimensionless.
The classical Yang-Mills Lagrangian describes the ensuing
locally conformally flat Einstein-Hilbert gravity which is coupled to
matter and with a nontrivial cosmological constant. The
matter multiplet consists of the massive vector field $J_\mu$,
the $O(3)$ $\sigma$-model described by $\vec n$, and the nonlinear
Grassmannian $\sigma$-model $G(4,2)$ which is described by the zweibeine
fields $e_\mu^a$. But in the reduced case (\ref{Lreduced}) the $O(3)$ matter
field $\vec n$ is decoupled on the classical level (it is classically trivial).

The reduced generally covariant gravitational Lagrangian which is relevant for
the present case where $\vec n$ is fixed to $\{0,0,1\}$ reads
\cite{FaddNiemi}
\beqa \la{Lgrav}
 L&=&
  \frac{1}{4}\sqrt{g} g^{\mu\nu} g^{\rho \sigma} {\mathcal F}_{\mu\rho} {\mathcal F}_{\nu\sigma}
+ \frac{\Delta^2}{12} R \sqrt{g}
+ \frac{1}{2}\Delta^2 \sqrt{g} g^{\mu\nu} J_\mu J_\nu
+ \Delta^2 \cdot \sqrt{g} \cdot g^{\mu\nu} g^{\lambda \eta}
 ({\bar {\mathcal D}}_{\mu \,\, \lambda}^{\,\, \sigma} \bar{\ce}_\sigma)
({\mathcal D}_{\nu \,\, \eta}^{\,\, \kappa} \ce^{}_\kappa)
 - \frac{3}{8} \Delta^4 \sqrt{g} \nn
\\  &&+ \frac12 \d_\mu (\rho \d_\mu \rho) - \frac12 \d_\nu\d_\mu (X_{\mu +} X_{\nu -})
\eeqa
Here $\ce_\mu$ is a covariant generalization of the
zweibein $e_\mu^a$, obtained with the vierbein $E^a_\mu$:
\beqa
  g_{\mu \nu} = \delta_{ab} E^a_\mu E^b_\mu\\
 \la{defce} \ce_\mu = E^a_\mu e_a.
\eeqa
Note that the original Yang-Mills indices are here
redefined {\it from Greek to Latin}. All variables are
also pulled to the curved space with the help of the vierbein, and
we refer to \cite{FaddNiemi} for details and for the full Lagrangian.
Naive comparison of this Lagrangian with the standard Lagrangian of gravity
leads to an estimate $\Delta = \sqrt{\frac3{4 \pi}} m_{Pl}$. But this is valid only under the
assumption that the coefficient in front of the scalar curvature term remains the same in the low-energy
effective action. The Yang-Mills scale $\Lambda_{YM}$ is a completely free parameter in this model,
because classical Yang-Mills Lagrangian is conformally invariant. The spacetime in which the latter is formulated
plays an auxiliary role and can be arbitrary rescaled, changing thereby $\Lambda_{YM}$. We expect that inclusion of other fields would
fix the $\Lambda_{YM}$ scale. The ``cosmological term'' in the Lagrangian should not be understood in a direct way,
because in the low-energy effective action it would be balanced by the effective potential for $\rho$.

Note the appearance of two boundary terms in the action.
Both are non-trivial for the instanton solution, and their
contribution can be localized to a small sphere that surrounds
the singular point $x=0$ that corresponds
to a gauge singularity.
The contribution to the action from these two boundary terms cancel each other,
they have the exactly opposite limits when $x \to 0$.

We now proceed to the spacetime geometry described by the instanton solution. We find for
the metric tensor (\ref{gdef})
\beq
 g = 1_{4 \! \times 4} \, (\rho/\Delta)^2 =H^2 \frac{\r^4}{(\r^2+x^2)^2 x^2} 1_{4 \! \times 4},
\eeq
where we have introduced a constant $ H^2 \equiv 8/(\Delta^2) $.

The scalar curvature of this metric is
\beq
 R(x)=-\frac{18\, r^4}{H^2 \r^4}+\frac{36\, r^2}{H^2 \r ^2}+\frac{6}{H^2} \scip r\equiv |x|
\eeq

In order to inspect the structure of the ensuing manifold, we first rewrite the conformally flat metric
in spherical coordinates:
\beq
g_{\mu\nu} dx^\mu dx^\nu = (\rho/\Delta)^2 \left((dr)^2 + r^2 (d\Omega_3)^2\right).
\eeq
After the change of variables
\beq \la{defz}
 r = \frac{\r}{\sqrt{e^{2 z} -1}}.
\eeq
 the metric becomes
\beq
 \left(\frac{ds}{H}\right)^2 = (dz)^2 + (1-e^{-2 z})^2 (d\Omega_3)^2.
\eeq

When $z \to 0$ which correspond to infinity in the initial coordinates, this metric becomes
\beq
 \left(\frac{ds}{H}\right)^2 = (dz)^2 + 4 z^2 (d\Omega_3)^2.
\eeq
Note the curvature singularity at the point $z \to 0$. In the vicinity of
this point the spacetime approaches ${\mathbb R}^4$ in spherical coordinates,
but the spherical surfaces $\mathbb S^3$ are doubly covered.

When  $z \to \infty$, which correspond to the location of the center of the
instanton in the original coordinates, the metric rapidly approaches
that of $\mathbb S^3 \times (0, \infty) $, and the scalar curvature
approaches the constant value $6/H^2$.
Consequently, asymptotically we have a cylinder of radius $H$
and the geometry is that of
an {\it infinite, asymptotically cylindrical cigar which is doubly wrapped at its beginning}
\footnote{I gratefully acknowledge the help
of Joe Minahan in clarifying the questions of geometry}.  Note that in terms of the original
coordinates the beginning of the cigar corresponds to the space-time infinity
and the infinite cylindrical end of the cigar corresponds to the center of the instanton.

We conclude that the instanton develops a finite radius hole at the
position of its center, with the resulting change
in the topology and boundary of the spacetime.

Note that the geometry also resolves the directional singularity that we have
in the Grassmannian matter field $e_\mu$; see \eq{einst}:
Since the cylindrical cigar has asymptotically constant radius, the Grassmannian vector
field becomes well defined. Explicitly, in the new coordinates
the covariant version of the Grassmannian zweibein field for $z \gg 1$ reads
\beq \la{ecenter}
 \ce \approx \frac{H e^{i \phi }}{\sqrt{2}} \left[\sin \delta  (i \cos \delta
+\cos \theta \sin \delta ) \sin
   \theta  \,\bm{d\phi} + (\cos \delta  \cos \theta -i \sin \delta ) \sin
   \delta  \,\bm{d\theta}+\sin \theta  \,\bm {d \delta} \right],
\eeq
where $\phi,\ \theta, \ \delta$ are the standard spherical coordinates on $\mathbb S^3$.
This expression has only
complex phase singularities which are irrelevant from the point of view of the
Grassmannian $\sigma$-model; These singularities only appear
in the $U(1)_I$ gauge degree of freedom.

As $z\to 0$ the zweibein is also non-singular and
tends to zero
\beq
\la{efar}
 \ce \approx  H z \sqrt{2} e^{i \phi }  \left[\sin \delta  (i \cos \delta
+\cos \theta \sin \delta ) \sin
   \theta  \,\bm{d\phi} + (\cos \delta  \cos \theta -i \sin \delta ) \sin
   \delta  \,\bm{d\theta}+\sin \theta  \,\bm {d \delta}\right].
\eeq

 We note that the present instanton geometry resembles the Euclidean
black hole (gravitational instanton) \cite{EguchiHanson} except that it does not obey the
asymptotic flatness conditions and its curvature tensor is
not self-dual.\footnote{there are other approaches, relating
instantons to Euclidean
 Schwarzschild
 black
hole in AdS5 space \cite{Rey}}. Indeed, there is no reason why
a {\it microscopic} theory of gravity should lead to asymptotic
flatness of Euclidean solutions,  the flat space should only
arise in the low energy limit.
Instead,
we propose that we have an example of a ``spacetime foam'', and the present
solution should be viewed as a ``quantum black hole''.


It is interesting to compare the boundary terms in (\ref{Lgrav})
with the standard boundary term for Einstein gravity for spacetimes
with boundaries \cite{HawkingGibbons}. There,
the boundary terms appear when second derivatives are removed.
Here, the action is already in the first-order form and instead we go to the
opposite direction,
\beq
\frac{1}{2}(\partial_\mu \rho)^2 \to \frac{\Delta^2}{12} R \sqrt{g} + \frac12 \d_\mu (\rho \d_\mu \rho)
\eeq
We conclude that the Gibbons-Hawking \cite{HawkingGibbons} boundary
term is $\frac12 \d_\mu (\rho \d_\mu \rho)$ and {\it vice versa},
\beq
   \frac12 (\rho \d_\mu \rho) = - \frac{\Delta^2}{6} g_{a b} \left(\nabla_{e^a} e^b \right)_\mu =
   - \frac{\Delta^2}{6} \Gamma^\nu_{\nu\mu},
\la{fundform}
\eeq
where $e^a$ are the basis vector fields and $\Gamma$ is a Christoffel connection.
The right-hand side of \eq{fundform} contracted with the normal vector to the
boundary surface is the trace of the second fundamental form
and it presents the generally covariant form for the boundary term.

Finally we only note that the meron solution \cite{Meron}
\beq \la{meron}
 A_\mu^a(x) =  \frac{ \bar\eta_{a\mu\nu} x^\nu}{x^2}
\eeq
also obeys the MAG condition, and it has the same type of singularity at its
center as the instanton. At infinity the geometry is quite different. Since the analysis is
straightforward we defer the details.

\section{Multi-Instanton case} \la{sec4}
We now proceed to study the (gravitational)
generally covariant interpretation of a multi-instanton
solution. However, due to technical difficulties and transparency of the presentation
we only consider explicitly the simplest case, described
by an an approximate superposition of two instantons. The general case
can be investigated by employing the ADHM construction \cite{ADHM}
but will not be studied here; we expect that our qualitative results
persist in the most general case.

Consider first the non-overlapping limit of two BPST instantons,
$y_i \gg \r_j $ where $y_i$ are the position 4-vectors
for the instantons.
In this limit the metric tensor is
\beq
g  =H^2 \left ( \frac{\r_1^2}{(\r_1^2+(x-y_1)^2)
|x-y_1|} +\frac{\r_2^2}{(\r_2^2+(x-y_2)^2) |x-y_2|} \right)^2 1_{4 \! \times 4},
\la{twoinst}
\eeq

Clearly, near the centers of the instantons the metric is again similar to a cylinder
$\mathbb S^3 \times \rm (0,\infty)$ of radius $H$.

We define $\r^2 = \r_1^2+\r_2^2$ and change the radial variable as $|x|=r=\frac{\r}{\sqrt{2 z}}$.
We also introduce $\mathbb S^3$ spherical coordinates for the remaining variables.
Notice that for the large $r$ the variable $z$ tends to the $z$
of the one-instanton case.
The metric now becomes
\beq
  ds^2  =H^2 \frac{\r^2}{(2z)^3} \left ( \frac{\r_1^2}{(\r_1^2+(x-y_1)^2) |x-y_1|} +\frac{\r_2^2}{(\r_2^2+(x-y_2)^2) |x-y_2|} \right)^2 \left( (2 z)^2 (d\Omega_3)^2 + (d z)^2\right) \\
\la{twoinstsph}
\eeq
where $(d\Omega_3)^2 =(d\delta)^2+(\sin \delta d\theta)^2 + (\sin \delta \sin \theta d\phi)^2$.

To the leading order $r \gg y$ (i.e. $z \ll 1$) this yields
\beq
\left( \frac{ds}{H} \right)^2 = \left( (2 z)^2 (d\Omega_3)^2 + (d z)^2\right).
\eeq
We here have again the doubly-covered sphere $\mathbb S^3$ of radius $z$
that characterizes the beginning of the cigar in the single instanton case.
As $z$ grows, these spheres become deformed and near each of the instanton
centers the metric diverges forming asymptotically an infinite  cylinder $\mathbb
S^3 \times (0,\infty)$ of radius $H$.
Note that the boundary of the space again corresponds to the centers of the
instantons, and the singularity in the Grassmannian vector field become
resolved in the same manner  as in the single instanton case.






\section{Monopole loops} \la{sec5}
 It was argued in \cite{Brower} and also seen directly from
 various numerical computations \cite{Teper} that  the multi-instanton solutions in
 the MAG have monopole loop singularities (see also \cite{Bruckmann2003} for
 an analytic example). It means that, really, an instanton ensemble in MAG is not a superposition
 of singular gauge instantons. Instead, a more general gauge transformation
with singularity along some contour (a.k.a. monopole
 loop) is needed to satisfy the global MAG condition (\ref{GlobalMAG}).
An explicit, analytic form of a monopole loop is not known
for a generic instanton ensemble. But for the dilute case it is known that
the loops are small,
circular and non-percolating, see \cite{Brower} for explicit expressions.

To construct the monopole loop one applies the  gauge transformation
\beqa
 \Omega^\dag &=& e^{i \gamma \tau_3/2} e^{i \beta \tau_2/2} e^{i \alpha \tau_3/2}
\nn
\eeqa
where
\beq
\alpha = \phi-\psi \\
\gamma=\pi - \phi - \psi
\eeq
and
\beqa
\beta = 2 \theta - \arctan \frac{u}{v+R} - \arctan \frac{u}{v-R}
\eeqa
to the anti-instanton in the singular gauge (or its conjugate $\Omega$ to the instanton).
Here the coordinates
\beq \bea{c}
      x_1 + i x_2 = u e^{i \phi} \\
       x_3+ i x_4  = v e^{i \psi}
     \ea
 \eeq
 are used.
This gives a monopole loop of radius $R$ that lies in the (3-4) plane and can be arbitrarily
reoriented by a global gauge transformation.

The corresponding value of $\rho^2$ is
\beq
 \rho^2=\frac{8 \left(\left(\r^2-R^2\right)^2 u^2+\left(\r^2+R^2\right)^2
   v^2\right)}{\left(\r^2+u^2+v^2\right)^2 \left( R^4+2 (u^2-v^2) R^2+\left(u^2+v^2\right)^2\right)}.
\eeq
  At $R=0$ the above equation reduces to \eq{rho_inst} for the singular
gauge instanton. The singularity here forms a ring of radius $R$.
Near this ring the condensate is
\beq \la{condring}
 \rho^2 = \frac{2}{z^2}+\frac{2 \left(\r^2-3 R^2\right) \cos (\eta )}{R \left(\r^2+R^2\right) z}+{\cal O} \left(z^0\right),
\eeq
 where we used natural coordinates for a ring:
 $z \cos \eta = (v-R) \ ; \ z \sin \eta = u $
 with $\eta \in [0,\pi)$.

The 3-vector $n$ is again trivial $n=\{0,0,1\}$,
so $$\bea{c}\psi_1 = \rho \\  \psi_2 = 0 \ea $$ and the zweibein field is simply related to the new gauge transformed potential:
\beq\
e_\mu = \frac{A^1_\mu + i A^2_\mu}{\sqrt2\rho}.
\eeq

 Keeping only the first term in (\ref{condring}) we analyze the qualitative behaviour of
 the corresponding geometry. In coordinates $\psi$ (angular position on the ring), $z,\ \eta, \ \phi$ (spherical
 coordinates around ring points) the metric associated with $\rho^2$ becomes
\beq
 ds^2 \approx \frac2{\Delta^2 z^2}\left(R^2\, d\psi^2 + dz^2 + z^2 \,d\eta^2 + z^2 \sin^2 \!\eta \; d\phi^2  \right).
\eeq
Changing the variable  $z=H e^{-y}$ results in:
\beq
 \left(\frac{ds}{H}\right)^2 \approx  \frac14 \left(e^{2 y} (R/H)^2\, d\psi^2 + dy^2 + d\eta^2 +  \sin^2 \!\eta \; d\phi^2\right).
\eeq
Thus the space is like infinite cylinders $(\eta,\phi,y) \in \mathbb S^2 \times (0,\infty)$ of
radius $H/2$ over each
point of the circle, parameterized by $\psi$. Cylinders over different points spread away from each other exponentially fast
as one moves along them (i.e. increases $y$). The scalar curvature tends to zero exponentially fast with increasing
of $y$.

When we are far from the monopole loop, the results obtained for the singular gauge are valid, thus the total
geometry looks like $\mathbb S^3$-cylinder that becomes split into infinite $\mathbb S^2$ cylinders.
In terms of the variable $z$ introduced for analysis of singular gauge instanton in (\ref{defz}),
the splitting occures at $z \approx \log(\r/R)$. The resulting cylinders with $\mathbb S^2$ profiles are fibered
over one of the $\mathbb S^1$ sections of $\mathbb S^3$, depending on the orientation of the monopole loop.
This splitting can be illustrated in lower dimensions in terms of gradual transition from a sphere to a torus.

 In the coordinates $\{y, \psi,\eta,\phi\}$, used above, the Grassmannian field $\ce$, defined in
 \eq{defce},  takes the form
\beq
{\ce} = \frac{i e^{i (\phi +\psi )} H (\r^2-R^2) \sin (\eta )}{2
   \left(\r^2+R^2\right)} \bm{ d\psi} + \frac{e^{i (\phi +\psi )} H}{2 } \bm{ d\eta} + \frac{i e^{i (\phi +\psi )} H \sin
   (\eta )}{2 } \bm{ d \phi} + {\cal O}(e^{-y})
\eeq

We again observe that the geometry resolves a directional singularity of the initial field $e_\mu$: the
new field $\ce$, defined on the space with metric (\ref{gdef}), is non-singular.
The singularity near the  monopole loop is mapped to a boundary surface at infinity.

\section{Discussion} \la{secDiscussion}
In summary, we have elaborated on the
observation \cite{FaddNiemi} that the Yang-Mills action in the MAG gauge
can be viewed as a locally conformally flat limit of a gravitational
theory that involves the Einstein action and a cosmological constant in interaction
with matter fields. In this approach the short distance Minkowski space
Yang-Mills theory then provides a mechanism
where the metric tensor of the locally conformally flat
spacetime becomes generated by the dimension-two
condensate in the gauge theory, by quantum fluctuations.

We have inspected the BPST instanton in this formalism.
We find that an instanton resembles a Euclidean black
hole configuration in the form of a spacetime that has
the shape of a doubly wrapped cigar. For a multi-instanton, we expect
a spacetime with several cigar-like cylinders,
growing from different
singular points as deformations of the space.
These cylinders asymptotically
approach $(0, a) \times \mathbb S^3$ and then split into
$(a, \infty) \times \mathbb S^2$ cylinders fibered over $\mathbb S^1 \subset \mathbb S^3$.  We propose that
this can be interpreted as a Euclidean ``spacetime foam'' solution, a
fluctuation that creates spacetime from ``nothing''.

\section*{Acknowledgements}
I thank Antti Niemi for inspiring this work, many comments
and for helping me by editing this article; I thank Joe Minahan,
Maxim Chernodub, Mikhail S. Volkov and Konstantin Zarembo for discussions and suggestions.
This work was partially supported by the Dmitri Zimin 'Dynasty' foundation,
RSGSS-1124.2003.2 ; RFFI project grant 06-02-16786, STINT Institutional Grant and
by a Grant from VR.


\begin{thebibliography}{99}
\bibitem{FaddNiemi}
  L.~D.~Faddeev and A.~J.~Niemi,
  arXiv:hep-th/0608111.
\bibitem{Gubarev2000}
  F.~V.~Gubarev and V.~I.~Zakharov,
  Phys.\ Lett.\ B {\bf 501}, 28 (2001)
  [arXiv:hep-ph/0010096].

\bibitem{Gubarev}
  F.~V.~Gubarev, L.~Stodolsky and V.~I.~Zakharov,
  Phys.\ Rev.\ Lett.\  {\bf 86}, 2220 (2001)
  [arXiv:hep-ph/0010057].
\bibitem{Verschelde}
  H.~Verschelde, K.~Knecht, K.~Van Acoleyen and M.~Vanderkelen,
  Phys.\ Lett.\ B {\bf 516}, 307 (2001)
  [arXiv:hep-th/0105018].
\bibitem{Kondo2001}
  K.~I.~Kondo,
  Phys.\ Lett.\ B {\bf 514}, 335 (2001)
  [arXiv:hep-th/0105299].
\bibitem{Brasil}
  D.~Dudal, J.~A.~Gracey, V.~E.~R.~Lemes, M.~S.~Sarandy, R.~F.~Sobreiro, S.~P.~Sorella and H.~Verschelde,
   ``An analytic study of the off-diagonal mass generation for Yang-Mills
  Phys.\ Rev.\ D {\bf 70}, 114038 (2004)
  [arXiv:hep-th/0406132].

\bibitem{CondensateReview}
  D.~Dudal, M.~A.~L.~Capri, J.~A.~Gracey, V.~E.~R.~Lemes, R.~F.~Sobreiro, S.~P.~Sorella and H.~Verschelde,
  arXiv:hep-th/0611114.
\bibitem{Boucaud}
  P.~Boucaud {\it et al.},
  Nucl.\ Phys.\ Proc.\ Suppl.\  {\bf 119}, 694 (2003)
  [arXiv:hep-ph/0209031].
  P.~Boucaud {\it et al.},
  Phys.\ Rev.\ D {\bf 66}, 034504 (2002)
  [arXiv:hep-ph/0203119].

\bibitem{Maxim}
 M.~N.~Chernodub,
  Phys.\ Lett.\ B {\bf 637}, 128 (2006)
  [arXiv:hep-th/0506107];
  JETP Lett.\  {\bf 83}, 268 (2006)
  [arXiv:hep-th/0507221].
\bibitem{BPST}
A. Belavin, A. Polyakov, A. Schwartz and Yu. Tyupkin, Phys. Lett. {\bf 59}, 85 (1975).
\bibitem{Gribov}
  V.~N.~Gribov,
  Nucl.\ Phys.\ B {\bf 139}, 1 (1978).


\bibitem{Bruckmann2000}
  F.~Bruckmann, T.~Heinzl, A.~Wipf and T.~Tok,
  Nucl.\ Phys.\ B {\bf 584}, 589 (2000)
  [arXiv:hep-th/0001175].
\bibitem{Brower}
  R.~C.~Brower, K.~N.~Orginos and C.~I.~Tan,
  Phys.\ Rev.\ D {\bf 55}, 6313 (1997)
  [arXiv:hep-th/9610101].
\bibitem{tHooft}
G.~'t~Hooft, Phys. Rev. D {\bf 14}, 3432 (1976).
\bibitem{EguchiHanson}
  T.~Eguchi and A.~J.~Hanson,
  Phys.\ Lett.\ B {\bf 74}, 249 (1978).


\bibitem{HawkingGibbons}
  G.~W.~Gibbons and S.~W.~Hawking,
  Phys.\ Rev.\ D {\bf 15}, 2752 (1977).
\bibitem{Meron}
V.~de Alfaro, S.~Fubini and G.~Furlan,
  Phys.\ Lett.\ B {\bf 65}, 163 (1976).


\bibitem{ADHM}
M.F.~Atiyah, V.G.~Drinfeld, N.J.~Hitchin and Yu.I.~Manin,
Phys.~Lett. A {\bf 65}, 185 (1978).

\bibitem{Teper}  A.~Hart and M.~Teper,
  Phys.\ Lett.\ B {\bf 371}, 261 (1996)
  [arXiv:hep-lat/9511016] ;   Phys.\ Rev.\ D {\bf 58}, 014504 (1998)
  [arXiv:hep-lat/9712003];
  \\V. Bornyakov, G. Schierholz DESY 96-069,
                  HLRZ 96-22, hep-lat/9605019 ;
                  G. Schierholz DESY 95-127 HLRZ 95-35; \\
 H. Markum, W.Sakuler, S. Thurner, Nucl. Phys B
               (Proc. Supl. LATTICE95 ), 254 (1995) (hep-lat/9510024);
               S. Thurner, M. Feurstein, H. Markum, W. Sakuler,
               Phys. Rev D54 3457(1996) \\
   M.~N.~Chernodub, F.~V.~Gubarev and M.~I.~Polikarpov,
  JETP Lett.\  {\bf 69}, 169 (1999)
  [arXiv:hep-lat/9801010];


  M.~Fukushima, H.~Suganuma and S.~Chiba,
Bruckmann2003
\bibitem{Bruckmann2003}
  F.~Bruckmann and D.~Hansen,
  Annals Phys.\  {\bf 308}, 201 (2003)
  [arXiv:hep-th/0305012].

\bibitem{Rey}
 S.-J. Rey and Y. Hikida,
JHEP 0608:051,2006 arXiv: hep-th/0507082;
 JHEP 0607:023,2006, arXiv: hep-th/0604102

\end{thebibliography}
\end{document}